\documentstyle[12pt]{article}
\begin{document}
\newcommand{\bea}{\begin{eqnarray}}
\newcommand{\eea}{\end{eqnarray}}
\renewcommand{\thefootnote}{\fnsymbol{footnote}}
\newcommand{\be}{\begin{equation}}      \newcommand{\ee}{\end{equation}}
\newcommand{\st}{\scriptsize}
\newcommand{\fs}{\footnotesize}
\vspace{1.3in}
\begin{center}
\baselineskip 0.3in

{\large \bf Eigenfunctions of spinless particles in a one-dimensional linear potential well}

\vspace {0.3in}
{ Nagalakshmi A. Rao}

\vspace{-0.2in}
{\it Department of Physics, Government Science College,}

\vspace{-0.2in}
{\it Bangalore-560001,Karnataka, India.}

\vspace{-0.2in}
{drnarao@gmail.com}
 
\vspace{0.2in}
{ B. A. Kagali }

\vspace{-0.2in}
{\it Department of Physics, Bangalore University,}

\vspace{-0.2in}
{\it Bangalore-560056, India.}

\vspace{-0.2in}
{bakagali@gmail.com}
\end{center}

\vspace{0.3in}
\hspace{2.3in}

\begin{center}
{\bf Abstract}
\end{center}
In the present paper, we work out the eigenfunctions  of spinless particles bound in a 
one-dimensional linear finite  range, attractive potential well, treating it as a 
time-like component of a four-vector. We show that the one-dimensional stationary Klein-Gordon
equation is reduced to a standard differential equation, whose solutions, consistent with the
boundary conditions, are the parabolic cylinder functions, which further reduce to
the well-known confluent hypergeometric functions.

\vspace{0.1in}

\vspace{0.6in}
\noindent
{\bf PACS(2008):} 03.65.Ge; 03.65 Pm.

\vspace{0.1in}
\noindent

\vspace{0.3in}
\noindent

\setcounter{section}{0}
\newpage
\indent
{\section{\large \bf Introduction}}
With ever-increasing applicability of relativistic wave equations in nuclear physics and other
 areas, the relativistic bound state solutions of the Klein-Gordon and Dirac equations for 
various potentials has drawn the attention of researchers. Setting apart the mathematical 
complexity and computational difficulty, there are certain unresolved questions in relativistic theory for treating a 
general potential. In fact, the way of incorporating a general potential is not unambigiously 
defined.

In literature, several authors [1 - 5] have addressed the bound states of  various kinds of 
linear potential. While Chiu [6] has examined the quarkonium systems with the 
regulated linear plus Coulomb potential in momentum space, Deloff [7] has used a
 semi-spectral Chebyshev method for numerically solving integral equations and has 
applied the same to the quarkonium bound state problem in momentum space. 

In recent years, Rao and Kagali [8 - 10] have analysed the bound states 
of spin-half and spin-zero particles in a screened Coulomb potential, having a linear
 behaviour near the origin and shown the existence of genuine bound states. Very recently, 
Rao and Kagali[11] have reported on the bound states of a non-relativistic particle in a finite,
short-range linearly rising potential, envisaged as a quark-confining potential. 
In the present paper, we explore the relativistic bound states of spinless particles 
in the one-dimensional linear potential by considering the celebrated Klein-Gordon equation.

\vspace{-0.1in} 
\indent
{\section{\large \bf The Klein-Gordon Equation with the Linear Potential}}
Since the early days of quantum mechanics, the relativistic investigation of various 
one-dimensional systems is considered to be important. The Klein-Gordon equation which 
essentially describes spin zero particles like the pions and kaons, is a second order wave
equation in space and time and indeed a Lorentz invariant. Presently, we explore the 
solutions of the stationary Klein-Gordon equation with the linear potential well, 
treating it as a time-like component of a four-vector.

\indent
The one-dimensional time-independent form of the Klein-Gordon equation for a free particle 
of mass $'m'$, is
\bea
\left[{d^{2}\over d{x}^{2}}+{{E}^{2}-{m}^{2}{c}^{4}\over {c}^{2}{\hbar
}^{2}}\right]\psi(x) = 0
\eea
For a general potential $V(x)$, treated as the fourth component of a Lorentz-vector,
 this equation takes the form[12]

\bea
\left[{d^{2}\over d{x}^{2}}+{{(E - V(x))}^{2}-{m}^{2}{c}^{4}\over {c}^{2}{\hbar
}^{2}}\right]\psi(x) = 0 . 
\eea

Thus for the potential $V(x)$, in the vector-coupling scheme, 
the above equation may be written as
\bea
\left[{d^{2}\over dx^{2}}+{E^{2}-2EV(x)+V^{2}(x)-m^{2}c^{4}\over c^{2}\hbar
^{2}}\right]\psi =0.
\eea

Interestingly, the above equation may be rewritten in the Schrodinger form, with an effective energy and 
effective potential as 
\begin{eqnarray}
\left[{d^{2}\over {dx}^{2}}+(E_{eff}-{V}_{eff})\right]\psi =0  
\end{eqnarray}

with $E_{\!\!eff}=${\Large ${E^{2}-{m}^{2}{c}^{4}\over {c}^{2}{\hbar }^{2}}$}
\ \ and \ \ ${V}_{\!\!eff}=${\Large ${2EV(x)-V^{2}(x)\over {c}^{2}{\hbar }^{2}}$}.

Since $E_{eff}$ and $V_{eff}$ are non-linear in E and V, some novel results may be expected.
As in the non-relativistic case, the allowed free-particle solution, outside the potential
 boundry, would yield 
\bea
\psi _{1}(x)={C_{1}}e^{\alpha x}\ \ \ -\infty <x\leq -a  
\eea 
\bea
\psi _{4}(x)={D_{1}}e^{-\alpha x}\ \ \ \ a\leq x<\infty ,
\eea
consistent with the requirement $\psi(x)$ vanishes as $\left|x\right|\rightarrow \infty $.
Here $\alpha ^{2}$ =- $E_{eff}$ is implied. 

To discuss the nature of the solution within the potential region, $-a<x<a,$ we 
 consider a simple linear rising, finite range potential of the form [11] 
\vspace{-0.1in}
\bea
V\left(x\right)=-{V_{0}\over a}\left(a-\left|x\right|\right)
\eea
in which the well depth $V_{0}$ and range $2a$ are positive and adjustable parameters.
Owing to its shape, this potential could also be called the trianglular potential well.
The linear, finite-ranged potential so constructed, serves as a good model to describe the
energy specrum of particles, both relativistically and non-relativistically. We
have recently reported that this potential has a rich set of solutions and can bind non-relativistic 
particles. Presently, we study the bound states of spin zero particles in this linear potential,
 treating it as a Lorentz vector.
 
Introducing the potential in Eqn.(3) and on simplification, we obtain, for $x>0$,
\bea
\left[{d^{2}\over dx^{2}}+{1\over c^{2}\hbar ^{2}}\left\{ {V_{0}^{2}\over
a^{2}}x^{2}-\left(2EV_{0}+2V_{0}^{2}\right){x\over
a}+\left(E+V_{0}\right)^{2}-m^{2}c^{4}\right\} \right]\psi =0
\eea
This equation may be written as
\bea
\left[{d^{2}\over dx^{2}}+{A\over a^{2}}\left(x^{2}\over a^{2}\right)+{B\over
a^{2}}\left(x\over a\right)+{C\over a^{2}}\right]\psi =0
\eea
where $A=\bar V_{0}^{2}$, $B=-2\bar E\bar V_{0}-2\bar V_{0}^{2}$ \ and \ $C=\left(\bar E+\bar V_{0}\right)^{2}-\bar m^{2}$.

Here $\bar V_{0} \ =$ \ {\Large ${V_{0 }\over \hbar c/a}$}, \ $\bar E\ =$\ {\Large ${E\over \left(\hbar
c/a\right)}$} \ and \ $\bar m\ =$\ {\Large ${mc^{2}\over \hbar c/a}$}.

It is trivial to note that $\bar V_{0},\ \bar E \ {\rm and } \ \bar m$  are all 
dimensionless quantities.

Defining a new variable
\vspace{-0.1in}
\begin{eqnarray*}
y={x\over a},
\end{eqnarray*}
Eqn.(9) transforms into a standard form [13]
\bea
{d^{2}\psi \over dy^{2}}+\left(Ay^{2}+By+C\right)\psi =0,
\eea
whose solutions are the well-known Parabolic Cylinder Functions.

Further, with the substitution  $z=2\sqrt {A}\left(y+{B\over 2A}\right),$ the above equation takes the form
\bea
4A{d^{2}\psi \over dz^{2}}+\left({z^{2}\over 4}-D\right)\psi =0
\eea
where $D=${\large ${B^{2}-4AC\over 4A}$} is implied.

Using the transformation $\rho ^{2}=${\large ${z^{2}\over \sqrt {4A}}$}, we obtain
\bea
{d^{2}\psi \over d\rho ^{2}}+\left({\rho ^{2}\over 4}-b\right)\psi =0
\eea
with $b=${\large ${B^{2}-4AC\over \left(4A\right)^{3\over 2}}$}.

It is straightforward to check that $b=${\large ${\bar m^{2}\over 2\bar V_{0}}$}$>0$, since both $\bar m$ and $\bar V_{0}$
are positive. The eigenfunctions of spinless particles, which are the solutions of Eqn.(12) 
may be written in terms of the confluent hypergeometric functions as
\bea
\psi =N\ exp\left(-{1\over 4}\rho ^{2}e^{i\pi \over 2}\right)M\left({-ib\over
2}+{1\over 4},{1\over 2},{1\over 2}\rho ^{2}e^{i\pi \over 2}\right)
\eea
Physically admissible solutions require finiteness and normalizability and as is evident 
from the above equation, we see that the wavefunction vanishes as $\left|x\right|\rightarrow \infty $, 
and thus being square integrable, represents genuine bound states. 

{\section{\large \bf Results and Discussion}}
In relativistic quantum mechanics, it is well-known that
a general potential can be introduced in the wave equation in two different ways following the {\it 
minimal coupling scheme}. While in vector coupling, the potential $V(x)$ is treated as the 
fourth component of a four vector field, in scalar coupling, it is added to the invariant mass.
Whereas the vector interaction is charge dependent and acts differently on particles and
antiparticles, the scalar interaction is independent of the charge of the particle
and has the same effect on both particles and antiparticles. Hence, it is interesting to 
study the quantum dynamics of relativistic particles for various interactions using 
different coupling schemes, with a view to decide on the appropriate 
prescription for a given potential. 

The linear potential, envisaged as a quark-confining potential, is central in particle physics.
Our investigation concerning the boundstates of spinless particle in the one-dimensional 
linear, finite-range potential, is seemingly interesting. It is trivial to note that the Klein 
Gordon equation can be reduced to a Schrodinger-like equation with an effective energy 
$E_{eff}$ and an effective potential $V_{eff}$. The illuminating relation between the Klein Gordon
equation and the Schrodinger equation with an equivalent energy dependent potential has a 
number of applications. If the potential is weak enough to ignore the $V{^2}$ term, the relativistic
formalism becomes equivalent to the non-relativistic formalism. More importantly, in situations
where the Klein-Gordon equation is not exactly solvable, the Schrodinger form of the KG equation 
sheds some light on the problem as it could be reduced to a solvable eigenvalue problem.

In the present work, we show that the one-dimensional 
Klein Gordon equation for the linear potential in the vector-coupling scheme is reduced to
a standard differential equation, whose solutions, 
consistent with the boundary condition are the parabolic cylinder functions, which on further
simplification yield the confluent hypergeometric functions. Apart from 
being elegant, the vector coupling prescription is particularly significant in the 
sense that it preserves gauge invariance. Such studies, 
apart from being pedagogical in nature, are potentially exciting and significant.

The linear potential well so described, has potential applications in electronics[14]. 
It would be interesting to study the Dirac bound states of such a linearly rising potential
of finite range, which would serve as a good model to describe the quarkonia.

\indent
{\large\bf Acknowledgements}

 This work was carried out under a grant and fellowship by the University Grants Commission. One
of the authors, (NAR) extends her thanks to Dr. N. Nagambika Devi, Commissioner for 
Collegiate Education in Karnataka for her endearing encouragement.

\newpage
\baselineskip 0.2in
{\large\bf References}
\newline
\newline
\newline $[1]$ Lichtenberg D B (1987) { Energy levels of quarkonia in potential models} 
{\it Int.J.Mod.Phys. A} {\bf 2} 1669
\newline
\newline $[2]$ Antippa A F and Phares A J (1978) {The linear potential: A solution in terms of 
combinatorics functions} {\it J.Math.Phys} {\bf 19} 308
\newline
\newline$[3]$ Antippa A F and Toan N K (1979) {The linear potential eigen energy equation I}
 {\it Can.J.Phys.} {\bf 57} 417
\newline
\newline$[4]$ Plante G and Antippa A F (2005) {Analytic solution of the Schrodinger equation for
 the Coulomb plus linear potential - The wave functions} {\it J.Math.Phys.} {\bf 46} 062108
\newline
\newline$[5]$ Antonio de Castro (2003) Bound states by a pseudoscalar Coulomb potential in one plus one
dimension arXiv:hepth/0303 175v2
\newline
\newline$[6]$ Chiu T W (1986) {Non-relativistic bound state problems in momentum space} 
{\it J.Phys.A Math.Gen.}{\bf 19} 2537
\newline
\newline$[7]$ Deloff A (2007) {Quarkonium bound state problem in momentum space revisted}
{\it Ann.Phys.} {\bf 322} 2315
\newline
\newline$[8]$ Nagalakshmi A Rao and Kagali B A  (2002) {Spinless particles in a screened Coulomb 
potential} {\it Phys.Lett. A} {\bf 296} 192
\newline
\newline$[9]$ Nagalakshmi A Rao and Kagali B A  (2002) {Dirac bound states in a one-dimensional 
scalar screened Coulomb potential} {\it Mod.Phys.Lett. A} {\bf 17} 2049
\newline
\newline$[10]$ Nagalakshmi A Rao and Kagali B A  (2002) {Bound states of Klein-Gordon particles
in scalar screened Coulomb potential} {\it Int.J.Mod.Phys.A} {\bf 17} 4793
\newline
\newline$[11]$ Nagalakshmi A Rao and Kagali B A (2008) {On the genuine bound states of a non-
relativistic particle in a linear finite range potential} {\it Elec. Jour. Th. Phys} {\bf 5}
169 
\newline
\newline$[12]$ strange P (1998) {\it Relativistic Quantum Mechanics} (Cambridge: 
Cambridge University Press)Chapter 9, Section 2.
\newline
\newline$[13]$ Abramowitz M and Stegun I A (1965) {\it Handbook of Mathematical Functions and
Formulas, Graphs and Mathematical Tables} (New York: Dover)
\newline
\newline$[14]$ Jasprit Singh (1997) {\it Quantum Mechanics -Fundamentals and Applications 
to Technology} (New York: A Wiley Interscience )
\newline

\end{document}